\documentclass[12pt]{article}

\usepackage{bbm}
\usepackage{amssymb,amsmath,amsthm}

\usepackage{graphicx}

\renewcommand{\arraystretch}{1.2}

\newcommand{\be}{\begin{equation}}
\newcommand{\ee}{\end{equation}}
\newcommand{\ba}{\begin{eqnarray}}
\newcommand{\ea}{\end{eqnarray}}

\newcommand{\bmat}{\begin{pmatrix}}
\newcommand{\emat}{\end{pmatrix}}

\newcommand{\mnu}{\mathcal{M}_\nu}

\newcommand{\s}{\hspace{0.5mm}}

\newcommand\ignore[1]{}

\allowdisplaybreaks

\begin{document}

\title{
\normalsize \hfill UWThPh-2011-31\\\hfill IFIC/11-47\\[12mm]
\LARGE The reactor mixing angle and CP violation with two texture zeros in the light of T2K
}

\author{
P.O.~Ludl\thanks{E-mail: patrick.ludl@univie.ac.at} $^{(a)}$,
S.~Morisi\thanks{E-mail: morisi@ific.uv.es} $^{(b)}$
and E.~Peinado\thanks{E-mail: epeinado@ific.uv.es} $^{(b)}$
\\*[3mm]
\small $^{(a)}$ University of Vienna, Faculty of Physics,\enspace Boltzmanngasse 5, A--1090 Vienna, Austria
\\*[3mm]
\small $^{(b)}$ AHEP Group, Institut de F\'{\i}sica Corpuscular -- C.S.I.C./Universitat de Val{\`e}ncia \\
\small Edificio Institutos de Paterna, Apt 22085, E--46071 Valencia, Spain
}

\date{24 January 2012}

\maketitle

\begin{abstract}
We reconsider the phenomenological implications of two texture zeros in symmetric neutrino mass matrices in the light of 
the recent T2K results for the reactor angle and the new global analysis which gives also best fit values for 
the Dirac CP phase $\delta$. The most important results of the analysis are: Among the viable cases classified by Frampton et al. only A$_1$ and A$_2$ predict
$\theta_{13}$ to be different from zero at 3$\sigma$. Furthermore these two cases are compatible only with a normal mass spectrum in the allowed region for the reactor angle.  At the best fit value A$_1$  and A$_2$ predict $0.024\ge\sin^2 \theta_{13}\ge0.012$ and $0.014\le\sin^2 \theta_{13}\le0.032$, respectively, where the bounds on the right and the left correspond to $\cos \delta=-1$ and $\cos \delta=1$, respectively. The cases B$_1$, B$_2$, B$_3$ and B$_4$ predict nearly maximal CP violation, i.e. $\cos \delta\approx 0$.
\end{abstract}

PACS-numbers: 14.60.-z,  
14.60.Pq,  
14.60.St,  
23.40.Bw.  

\section{Introduction}

Recently the T2K Collaboration~\cite{T2K} gave hints for a nonzero reactor angle, and also the results of the
MINOS Collaboration~\cite{MINOS} point towards the same direction. The global fits of neutrino oscillation 
experiments give\footnote{Throughout this work the abbreviations NS and IS will stand for normal and inverted neutrino mass spectrum, respectively.}
\begin{equation}
\begin{array}{lr}
\begin{array}{l}
0.001 \le \sin^2\theta_{13}\le 0.035~~~(\mbox{NS}),\\
0.001 \le \sin^2\theta_{13}\le 0.039~~~(\mbox{IS}),
\end{array} &\cite{Schwetz:2011zk} \\\\
\begin{array}{c}
0.005\le \sin^2\theta_{13} \le 0.050,
\end{array}& \cite{Fogli:2011qn}
\end{array}
\end{equation}
and the best fit values are $\sin^2\theta_{13}=0.013$ and $\sin^2\theta_{13}=0.025$, respectively. 
In particular the global analysis by Schwetz, Tortola and Valle\,\cite{Schwetz:2011zk} gives 
a weak hint for a nonvanishing CP violating phase, namely (at the best fit point)
\begin{eqnarray}
&&\sin^2\theta_{13}=0.013,\qquad \delta=-0.61\,\pi \quad(\mbox{NS}),\\
&&\sin^2\theta_{13}=0.016,\qquad \delta=-0.41\,\pi \quad(\mbox{IS}).
\end{eqnarray}

A lot of papers have been proposed recently in order to reproduce such a large value of the reactor mixing angle\,\cite{allmodels}.
Already before the recent T2K data there have been models based on discrete flavor symmetries  which predict a large reactor mixing angle---for an incomplete list see Ref.~\cite{Hirsch:2007kh}, and  for a classification of models with flavor symmetries classified by their predictions for the reactor angle see~\cite{Albright:2009cn}.

Here we reconsider the interesting case of Majorana neutrino mass matrices with two texture zeros in the basis where the charged lepton mass matrix is diagonal, which has been
extensively studied in the past. Our aim is to point out the phenomenological implication of such 
textures in the light of the T2K results\footnote{While we were finishing this work two papers treating the same problem were published in~\cite{Fritzsch:2011qv,Kumar:2011vf}.}. 

It was shown in~\cite{FGM, Xing} and~\cite{Fritzsch:2011qv} that, in the basis where the charged lepton mass matrix is diagonal, there are seven types of two texture zeros in symmetric neutrino mass matrices compatible with the experimental data on neutrino oscillations. In this work we want to analyze the correlation between the CP violating phase $\delta$ and the reactor mixing angle $\theta_{13}$ in the framework of these two texture zeros.

Another interesting possibility is to place texture zeros in the inverted neutrino mass matrix--see e.g.~\cite{Lavoura}. The implications of this type of two texture zeros on the reactor mixing angle and CP violation have been studied in~\cite{Verma}.

\section{Two texture zeros}

In the basis where the charged lepton mass matrix is diagonal, the lepton mixing matrix $U$ and the Majorana neutrino mass matrix $\mathcal{M}_{\nu}$ are related via
	\be\label{Mnu}
	\mathcal{M}_{\nu}=U^{\ast}\s\mathrm{diag}(m_1,m_2,m_3)\s U^{\dagger}.
	\ee
The standard parameterization~\cite{PDG} for $U$ is given by\footnote{The parameterization used here is a re-writing of the symmetrical parameterization proposed in~\cite{Schechter:1980gr}.}
	\be\label{Uparam}
	U=e^{i\hat{\alpha}}Ve^{i\hat{\sigma}},
	\ee
where
	\be
	V = \left( \begin{array}{ccc}
	c_{13} c_{12} &
	c_{13} s_{12} &
	s_{13} e^{-i \delta} \\
	- c_{23} s_{12} - s_{23} s_{13} c_{12} e^{i \delta} &
	c_{23} c_{12} - s_{23} s_{13} s_{12} e^{i \delta} &
	s_{23} c_{13} \\
	s_{23} s_{12} - c_{23} s_{13} c_{12} e^{i \delta} &
	-s_{23} c_{12} - c_{23} s_{13} s_{12} e^{i \delta} &
	c_{23} c_{13} \end{array} \right),
	\ee
$\hat{\alpha}=\mathrm{diag}(\alpha_1,\alpha_2,\alpha_3)$ and $\hat{\sigma}=\mathrm{diag}(\sigma_1,\sigma_2,\sigma_3)$. $c_{ij}=\mathrm{cos}\s\theta_{ij}$, $s_{ij}=\mathrm{sin}\s\theta_{ij}$ with $\theta_{ij}\in [0,\pi/2]$. $\delta\in[0,2\pi)$ is the CP violating phase and the $\sigma_i\in[0,2\pi)$ are the Majorana phases, which are not measurable in oscillation experiments. The phases $\alpha_i$ are irrelevant for neutrino oscillations and will play no role in our analysis, as we will see in the following. Inserting (\ref{Uparam}) into (\ref{Mnu}) we obtain
	\be
	(\mathcal{M}_{\nu})_{ij}^{\ast}=\sum_{k} m_k U_{ik}U_{jk}=\sum_k m_k e^{2i\sigma_k}V_{ik}V_{jk}e^{i(\alpha_i+\alpha_j)}.
	\ee
Placing a texture zero in the neutrino mass matrix corresponds to the condition
	\be
	(\mathcal{M}_{\nu})_{ij}=0 \quad (\Leftrightarrow (\mathcal{M}_{\nu})_{ij}^{\ast}=0 ) 
	\ee
for some indices $(i,j)$. Defining $\mu_k:=m_k e^{2i\sigma_k}$ and dividing by $e^{i(\alpha_i+\alpha_j)}$ we arrive at
	\be
	\sum_k \mu_k V_{ik}V_{jk}=0.
	\ee
The assumption of two texture zeros can thus be described by the two equations
	\be\label{system}
	\sum_{k}\mu_{k} V_{ak}V_{bk}=0,\quad \sum_k \mu_k V_{ck}V_{dk}=0.
	\ee
The viable cases of two texture zeros given in~\cite{FGM}  and the corresponding parameters $(a,b,c,d)$ can be found in table \ref{tab}. 
\begin{table}[h!]
\renewcommand{\arraystretch}{1.2}
\begin{center}                   
\begin{tabular}{c|c|c}             
case & texture zeros & (a,b,c,d) \\
\hline   
A$_1$ & $(\mnu)_{ee} = (\mnu)_{e\mu} = 0$ & (1,1,1,2) \\
A$_2$ & $(\mnu)_{ee} = (\mnu)_{e\tau} = 0$ & (1,1,1,3) \\
B$_1$ & $(\mnu)_{\mu\mu} = (\mnu)_{e\tau} = 0$ & (2,2,1,3)\\
B$_2$ & $(\mnu)_{\tau\tau} = (\mnu)_{e\mu} = 0$ & (3,3,1,2) \\
B$_3$ & $(\mnu)_{\mu\mu} = (\mnu)_{e\mu} = 0$ & (2,2,1,2) \\  
B$_4$ & $(\mnu)_{\tau\tau} = (\mnu)_{e\tau} = 0$ & (3,3,1,3) \\
C     & $(\mnu)_{\mu\mu} = (\mnu)_{\tau\tau} = 0$ & (2,2,3,3) 
\end{tabular}                                      
\end{center}                                       
\caption{The viable cases in the framework of two texture zeros in the Majorana neutrino mass matrix $\mnu$ and a diagonal charged-lepton mass matrix $\mathcal{M}_\ell$~\cite{FGM}. \label{tab}} 
\end{table}

\section{General remarks}

The system~(\ref{system}) is equivalent to
	\be
	\bmat
	V_{a1}V_{b1} & V_{a2}V_{b2}\\
	V_{c1}V_{d1} & V_{c2}V_{d2}\\
	\emat
	\bmat
	\mu_1\\
	\mu_2
	\emat=
	-\mu_3
	\bmat
	V_{a3}V_{b3}\\
	V_{c3}V_{d3}
	\emat.
	\ee
The set of solutions of this system of linear equations depends on the determinant
	\be
	D_{abcd}:=\mathrm{det}\bmat
	V_{a1}V_{b1} & V_{a2}V_{b2}\\
	V_{c1}V_{d1} & V_{c2}V_{d2}\\
	\emat=V_{a1}V_{b1}V_{c2}V_{d2}-V_{a2}V_{b2}V_{c1}V_{d1}.
	\ee
For $D_{abcd}\neq 0$ we find
	\be\label{mu1mu2}
	\bmat
	\mu_1\\
	\mu_2
	\emat=
	-\frac{\mu_3}{D_{abcd}}
	\bmat
	V_{c2}V_{d2} & -V_{a2}V_{b2}\\
	-V_{c1}V_{d1} & V_{a1}V_{b1}\\
	\emat
	\bmat
	V_{a3}V_{b3}\\
	V_{c3}V_{d3}
	\emat.
	\ee
Since at least two neutrino masses must be nonzero, the above equation implies that the lightest neutrino mass is different from zero.\footnote{
In general a normal (inverted) neutrino mass spectrum allows $\mu_1=0$ ($\mu_3=0$). However, one can verify that within the experimental 3$\sigma$-range (\ref{mu1mu2}) implies $\mu_1=0\Leftrightarrow \mu_3=0$ for all types of two texture zeros we will study in this work. Thus the lightest neutrino mass must be nonzero.}
Thus we are allowed to divide by $\mu_3$ and we can easily calculate
	\be\label{fraction}
	r:=\frac{\Delta m_{21}^2}{\Delta m_{31}^2}=\frac{\frac{m_2^2}{m_3^2}-\frac{m_1^2}{m_3^2}}{1-\frac{m_1^2}{m_3^2}}=\frac{\left\vert\frac{\mu_2}{\mu_3}\right\vert^2 - \left\vert\frac{\mu_1}{\mu_3}\right\vert^2 }{1-\left\vert\frac{\mu_1}{\mu_3}\right\vert^2}.
	\ee
Inserting (\ref{mu1mu2}) into (\ref{fraction}) we find an equation which relates the six quantities
	\[
	\Delta m_{21}^2,\,\Delta m_{31}^2,\, \theta_{12},\,\theta_{23},\,\theta_{13} \mbox{ and } \delta.
	\]
Fixing the mass squared differences and the two mixing angles $\theta_{12}$ and $\theta_{23}$ (e.g. to their best fit values or their $n\sigma$-ranges), we obtain a relation between the reactor mixing angle $\theta_{13}$ and $\delta$. Note that in this way one can eliminate the unknown absolute neutrino mass scale. This approach has been previously used in~\cite{Meloni}.

The main question we have to answer before beginning our analysis is whether the determinant $D_{abcd}$ can become zero for the seven different cases within the experimental limits. The first issue we notice is that all entries of the $2\times 2-$matrix
	\be
	\bmat
	V_{a1}V_{b1} & V_{a2}V_{b2}\\
	V_{c1}V_{d1} & V_{c2}V_{d2}\\
	\emat
	\ee
are nonzero (by experiment). Thus $D_{abcd}=0$ implies
	\be
	\frac{V_{a1}V_{b1}}{V_{c1}V_{d1}}=\frac{V_{a2}V_{b2}}{V_{c2}V_{d2}}.
	\ee
Using the experimentally known fact that the absolute values of all elements of the second column of $V$ are of the same order of magnitude, we find
	\be\label{detcondition}
	\left\vert\frac{V_{a1}V_{b1}}{V_{c1}V_{d1}}\right\vert=\left\vert\frac{V_{a2}V_{b2}}{V_{c2}V_{d2}}\right\vert\simeq 1\Rightarrow \vert V_{a1}V_{b1}\vert\simeq\vert V_{c1}V_{d1}\vert.
	\ee
From
	\be
	\vert V_{i1}\vert\simeq (2/\sqrt{6},\, 1/\sqrt{6},\, 1/\sqrt{6})^{\mathrm{T}}
	\ee
one easily finds that the only case allowing (\ref{detcondition}) is C. Therefore for the cases A$_1$ to B$_4$ we can assume $D_{abcd}\neq 0$ and use equation (\ref{fraction}) for our analysis. 

Let us now turn to case C. In~\cite{LG} it has been shown that for $\theta_{13}=0$ the determinant $D_{2233}$ becomes zero and the system~(\ref{system}) is therefore singular in this case.
Inversely assuming $D_{2233}=0$, we can proceed as follows.
Defining $\epsilon=s_{13}e^{i\delta}$ we find
	\be
	D_{2233} = \epsilon\,\left( \frac{1}{2}\,\mathrm{sin}(2\theta_{12})\mathrm{sin}(2\theta_{23})(1+\epsilon^2)-\epsilon\,\mathrm{cos}(2\theta_{12})\mathrm{cos}(2\theta_{23})\right).
	\ee
Thus $D_{2233}$ can be zero only for $\epsilon=0$ or 
\begin{equation}
\tan 2 \theta_{12}\,\tan 2\theta_{23}= \frac{2\epsilon}{1+\epsilon^2} .
\end{equation}
For $0\leq s_{13}^2\leq0.05$ we find
	\be
	\left\vert\frac{2\epsilon}{1+\epsilon^2}\right\vert=\frac{2 s_{13}}{\vert1+s_{13}^2\exp(2i\delta)\vert}\leq\frac{2 s_{13}}{1-s_{13}^2}<0.48\,.
	\ee
Using the $3\sigma-$ranges provided in~\cite{Schwetz:2011zk} one easily finds that at $3\sigma$
	\be
	\tan 2 \theta_{12}\,\tan 2\theta_{23}>8.55,
	\ee
which implies that $s_{13}=0$ is indeed the only possibility for $D_{2233}$ to become 0 at $3\sigma$. Since we are not interested in the limit $s_{13}\rightarrow 0$, we can use~(\ref{system}) and~(\ref{fraction}) also to analyze case C.

\subsection*{Analysis of the relation between $\sin^2\theta_{13}$ and $\cos\delta$}
It turns out that for all texture zeros studied in this work $r$ (see equation (\ref{fraction})) can be expressed as a rational function of at most cubic polynomials in $\mathrm{cos}\,\delta$, i.e.
	\be\label{equationcosdelta0}
	r=\frac{p(\cos\delta)}{q(\cos\delta)},	
	\ee
where $p$ and $q$ are polynomials of order at most 3. Thus we find
	\be\label{equationcosdelta}
	r\,q(\cos\delta)-p(\cos\delta)=0,
	\ee
which is an equation of at most third order in $\cos\delta$.
 Thus the dependence of $\mathrm{cos}\,\delta$ on the mixing angles can be computed \textit{exactly}. Note that~(\ref{equationcosdelta}) will in general have more solutions than~(\ref{equationcosdelta0}), because we have multiplied by $q(\cos\delta)$. In fact we have the additional solution
	\be
	q(\cos\delta)=p(\cos\delta)=0,
	\ee
which corresponds to the limit $\Delta m_{ij}^2/m_3^2\rightarrow 0$ (see equation (\ref{fraction})), i.e. a \textit{quasi-degenerate} neutrino mass spectrum.

For the cases A$_1$ and A$_2$ (\ref{equationcosdelta}) is linear in $\cos\delta$. B$_1$, B$_2$ and C lead to quadratic equations and B$_3$ and B$_4$ yield cubic equations for $\cos\delta$, respectively. We used Mathematica to obtain the coefficients of~(\ref{equationcosdelta}). The well-known formulae for the general solutions of quadratic and cubic equations were implemented in \textit{C}-programs which, scanning over the experimentally allowed ranges for $r$, $\theta_{12}$ and $\theta_{23}$, allowed us to plot $\mathrm{sin}^2\theta_{13}$ versus $\mathrm{cos}\,\delta$. $\mathrm{sin}^2\theta_{13}$ was varied between $0$ and $0.05$ and for all other experimentally accessible quantities we used the values obtained from the newest global fit~\cite{Schwetz:2011zk} including already the new T2K data~\cite{T2K}.
Our numerical analysis consists of the following steps:
	\begin{itemize}
	 \item[$-$] Input: $\sin^2\theta_{13}$, $\sin\theta_{12}$, $\sin\theta_{23}$, $r=\Delta m_{21}^2/\Delta m_{31}^2$ (best fit values or $n\sigma-$range ($n=1,2,3$)). The range of $\sin^2\theta_{13}$ (0 to 0.05) is divided into 600 steps. The ranges of $r,\,\sin\theta_{23}$ and $\sin\theta_{12}$ are divided into steps of equal length in such a way that the corresponding $1\sigma-$ranges are divided into 40 steps. Thus e.g. the computation for the 1$\sigma$-range alone consists of $600\times40^3=3.84\times10^7$ individual calculation cycles.
	 \item[$-$] Equation (\ref{equationcosdelta}) is solved for $\cos\delta$ (there can be up to three solutions).
	 \item[$-$] Only the real solutions $\in[-1,1]$ are processed further, the others are discarded.
	 \item[$-$] Now we want to insert the remaining solutions for $\cos\delta$ into~(\ref{mu1mu2}) to calculate the mass ratios 
				\be\label{massratios}
				\frac{m_i}{m_j}=\left\vert \frac{\mu_i}{\mu_j}\right\vert.
				\ee
In order to do that we have to calculate $e^{i\delta}$ from $\cos\delta$. There are two solutions to this problem, namely
				\be
				e^{i\delta}:=\cos\delta\pm i\,\sqrt{1-\cos^2\delta},
				\ee
but since they are complex conjugates of each other the mass ratios~(\ref{massratios}) do not depend on the choice of the solution. 
	\item[$-$] Finally the program checks whether the following inequalities are fulfilled.
			\be\label{massinequality}
			\begin{split}
			 & \frac{m_1}{m_3}<1,\,\frac{m_2}{m_3}<1,\,\frac{m_1}{m_2}<1\quad\mbox{(NS)},\\
			 & \frac{m_1}{m_3}>1,\,\frac{m_2}{m_3}>1,\,\frac{m_1}{m_2}<1\quad\mbox{(IS)}.
			\end{split}
			\ee
If they are fulfilled the data point ($\cos\delta$, $\sin^2\theta_{13}$) is stored.
	\item[$-$] Finally, when the whole parameter range has been scanned, all stored data points are plotted\footnote{In order to create plots of a suitable size (in terms of disk space) we constructed a lattice dividing the range of $\sin^2\theta_{13}$ into 600 and the range of $\cos\delta$ ($-1$ to $1$) into 800 points. Data points falling into the same part of the lattice were plotted only once.}.
	\end{itemize}

It turns out that a simple scan over the allowed $n\sigma$-ranges for the parameters as described above works very well for all cases of types A and B. However, for case C (NS) the sizes of the steps chosen in our systematic scan are just too large in order to obtain enough data points to produce good and reliable results. The reason for this issue is that C (NS) implies $\theta_{23}\approx 45^{\circ}$~\cite{LG}, so one would need an enormously high resolution in the scan over the $n\sigma$-ranges of $\sin\theta_{23}$ to produce reliable results. Thus we have to analyze C (NS) in a different way. A good method to deal with case C (NS) is to assign the input parameters $\sin^2\theta_{13}$, $\sin\theta_{12}$, $\sin\theta_{23}$, $r=\Delta m_{21}^2/\Delta m_{31}^2$ \textit{random} values in their $n\sigma$-ranges, rather than varying them step by step. In this way one obtains a so-called \textit{scatter plot}. Since also case C (IS) shows some hints of problems using a systematic scan, we also did a scatter plot for this case. The number of random points~$(\sin^2\theta_{13},\,\sin\theta_{12},\,\sin\theta_{23},\, r)$ we used was $10^9$ for each of the $n\sigma$-ranges ($n=1,2,3$).

\section{Results}
\begin{figure}
\begin{center}
\includegraphics[scale=0.50,angle=-90]{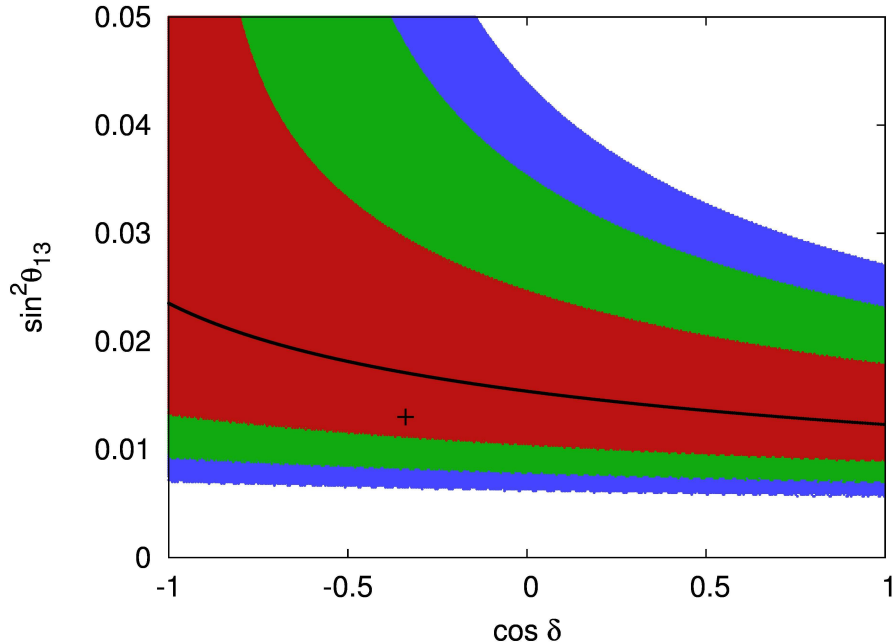}
\caption{The relation between $\sin^2\theta_{13}$ and $\cos\delta$ for case A$_1$ (normal spectrum). For description of the colors see the text.}
\label{a1normal}
\end{center}
\end{figure}

\begin{figure}
\begin{center}
\includegraphics[scale=0.50,angle=-90]{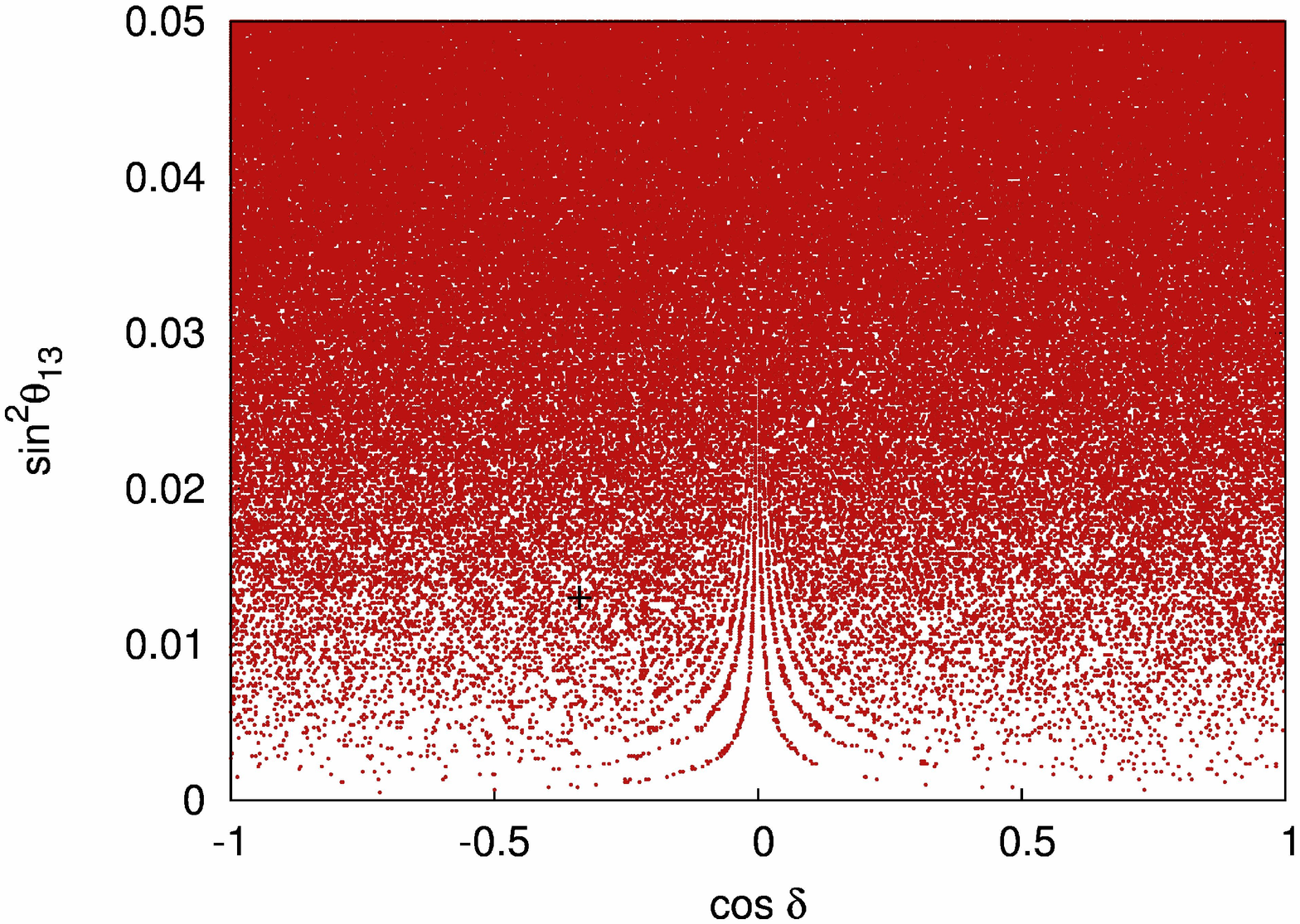}
\caption{The relation between $\sin^2\theta_{13}$ and $\cos\delta$ for case C (normal spectrum).\label{cnormal}}
\end{center}
\end{figure}

\begin{figure}
\begin{center}
\includegraphics[scale=0.50,angle=-90]{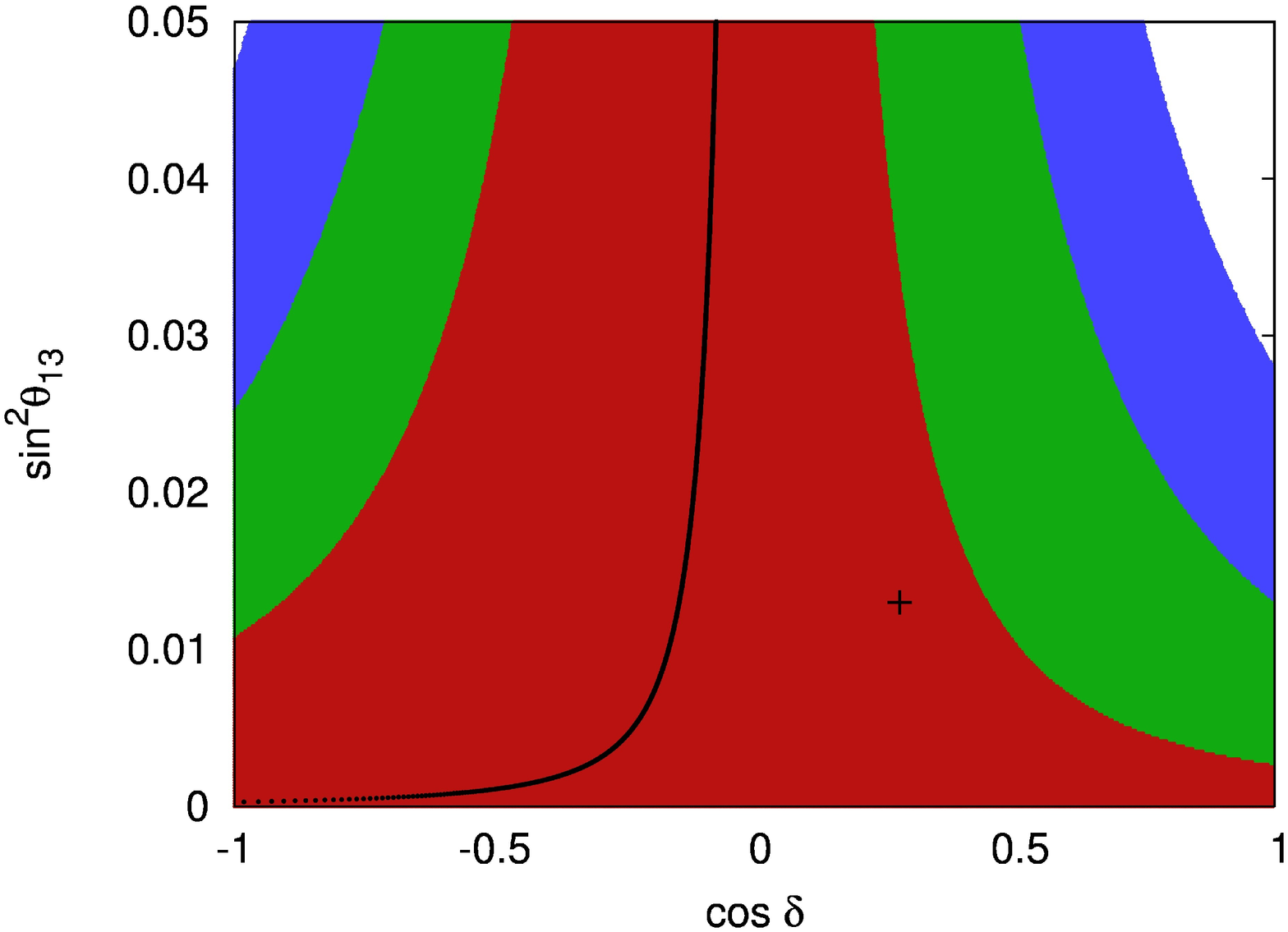}
\caption{The relation between $\sin^2\theta_{13}$ and $\cos\delta$ for case C (inverted spectrum).\label{cinverted}}
\end{center}
\end{figure}

We will now present the results of our numerical analysis. As already explained we have produced plots of $\sin^2\theta_{13}$ versus $\cos\delta$ (see figures \ref{a1normal}--\ref{cinverted}). The color code is the same for all plots:
	\begin{itemize}
	\item The best fit value for the point $(\cos\delta, \sin^2\theta_{13})$ according to the global fit~\cite{Schwetz:2011zk} is indicated by a black cross.	
	\item The best fit values for $\sin^2\theta_{13}$ as a function of $\cos\delta$ according to our analysis are indicated
		by a black line.
	\item The $n\sigma-$regions are shown as colored areas in the plots (red=1$\sigma$,\enspace green=2$\sigma$ and blue=3$\sigma$).
	\end{itemize}
The cases A$_1$ and A$_2$ are incompatible with an inverted spectrum if the reactor angle is varied within the $3\sigma$-range $0\le s^2_{13}\le0.05$. Assuming a normal neutrino mass spectrum A$_1$ and A$_2$ predict
$\theta_{13}$ to be different from zero at 3$\sigma$. For the best fit values for the observables, namely $\theta_{12}$, $\theta_{23}$ and $r$, we predict $0.024\ge\sin^2 \theta_{13}\ge0.012$ corresponding to the bounds $-1\le\cos\delta\le1$ for the case A$_1$ and $0.014\le\sin^2 \theta_{13}\le0.032$ corresponding to $-1\le\cos\delta\le1$ for the case A$_2$.

The cases B$_1$, B$_2$, B$_3$ and B$_4$ predict the Dirac CP phase to be close to maximal, i.e. $\cos \delta\approx 0$. Note furthermore that the cases B$_1$ (NS), B$_2$ (IS), B$_3$ (NS) and B$_4$ (IS) are incompatible with $\theta_{23}>45^{\circ}$.\footnote{This is in accordance with the best fit results for $\sin^2\theta_{23}$ given in~\cite{GrimusLudl} for B$_3$ and B$_4$.} Therefore, for these cases, the plots do not show the black ``best fit line'' (the best fit value for $\sin^2\theta_{23}$ given in~\cite{Schwetz:2011zk} is $0.52$ which corresponds to an atmospheric mixing angle larger than 45$^{\circ}$). The generic predictions we found for the cases of type B concerning the atmospheric angle are shown in table~\ref{Batmos}.
\begin{table}[h!]
\renewcommand{\arraystretch}{1.2}
\begin{center}                   
\begin{tabular}{|c|c|c|}             
\hline   case & NS & IS \\
\hline   
B$_1$ & $\theta_{23}\le 45^{\circ}$ & $\theta_{23} \ge45^{\circ}$ \\
B$_2$ & $\theta_{23}\ge 45^{\circ}$ & $\theta_{23}\le 45^{\circ}$ \\
B$_3$ & $\theta_{23}\le 45^{\circ}$ & $\theta_{23}\ge 45^{\circ}$ \\
B$_4$ & $\theta_{23}\ge 45^{\circ}$ & $\theta_{23}\le 45^{\circ}$ \\\hline   
\end{tabular}                                      
\end{center}                                       
\caption{Inequalities for the atmospheric mixing angle for the cases of type B. \label{Batmos}} 
\end{table}

\begin{figure}
\begin{center}
\includegraphics[scale=0.50,angle=-90]{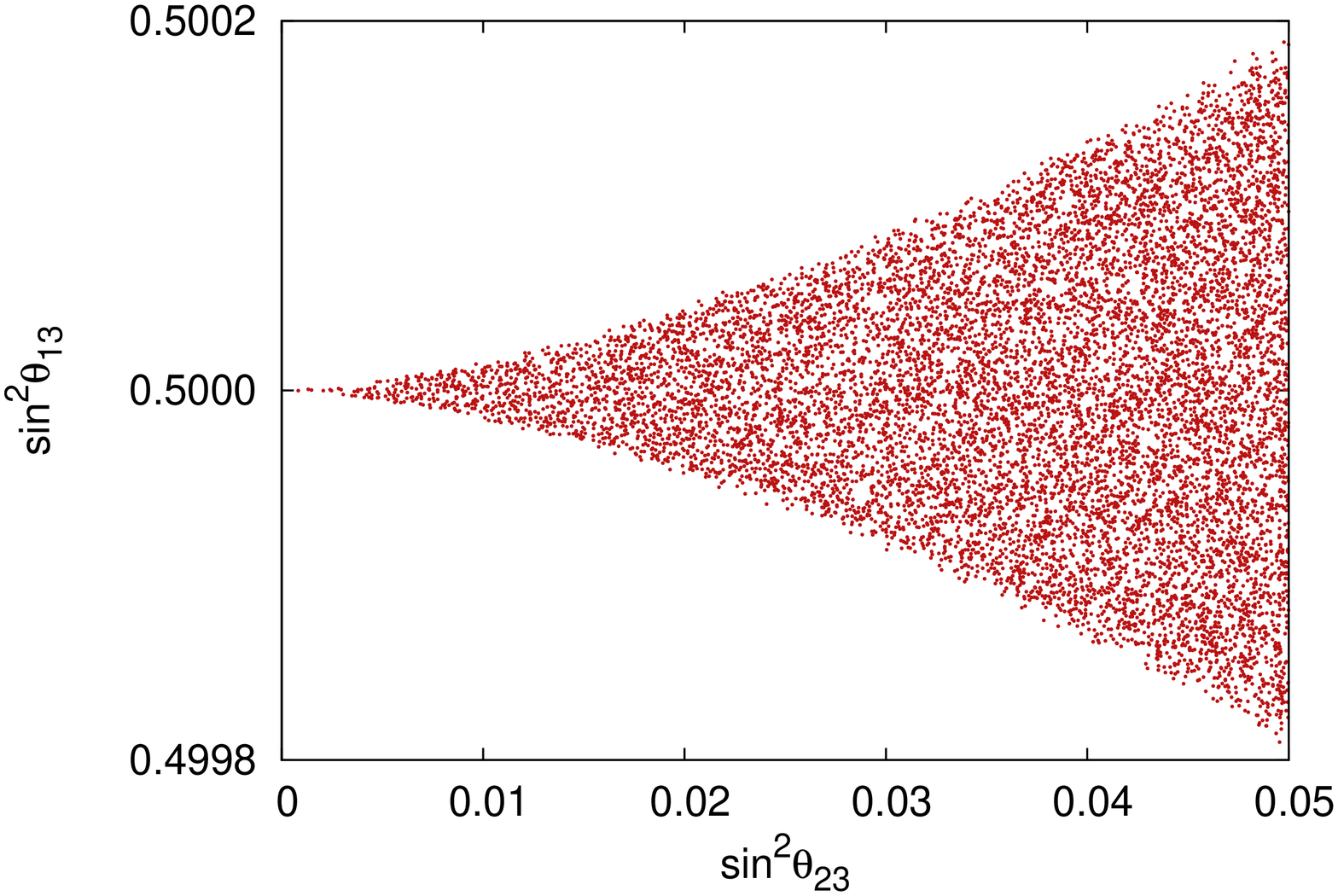}
\caption{The relation between $\sin^2\theta_{13}$ and $\sin^2 \theta_{23}$ for case C (NS) (scatter plot for the $1\sigma$-range with $10^7$ random points).\label{cnormals13s23}}
\end{center}
\end{figure}

For case C and an inverted spectrum we do not find a strong correlation between $\sin^2\theta_{13}$ and $\cos\delta$. However, from figure \ref{cinverted} we can see that the best fit value for the point $(\cos\delta, \sin^2\theta_{12})$ lies in the $1\sigma$-region of the plot. For case C and a normal spectrum there is no correlation between the Dirac CP phase and the reactor angle (see figure \ref{cnormal}), but, as was pointed out by Grimus and Lavoura~\cite{LG}, atmospheric neutrino mixing is close to maximal. As shown in figure~\ref{cnormals13s23} there is a correlation between the reactor angle and the atmospheric angle, but the deviation from the maximal value of the atmospheric angle is negligible.

\begin{figure}
\begin{center}
\includegraphics[scale=0.50,angle=-90]{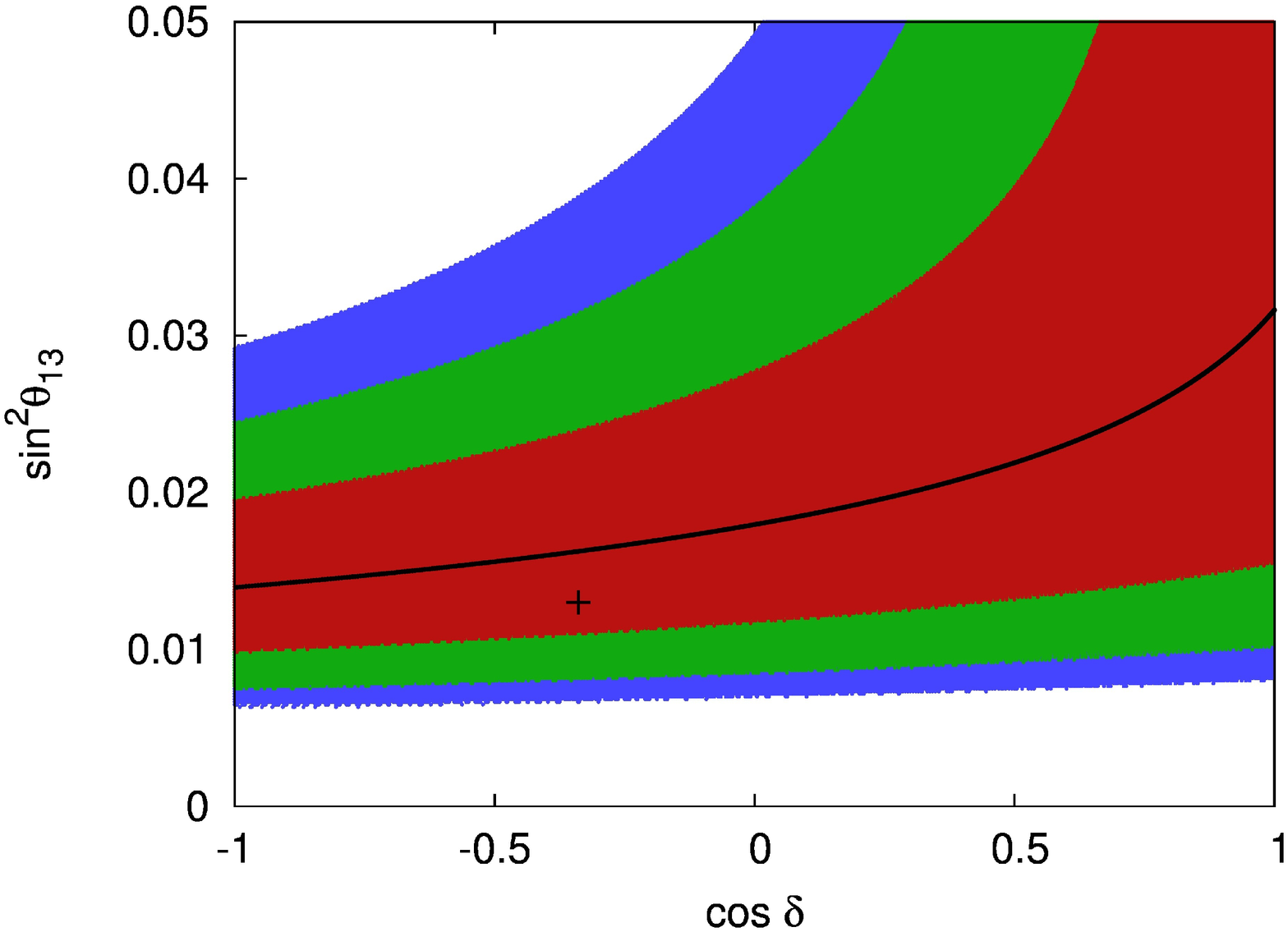}
\caption{The relation between $\sin^2\theta_{13}$ and $\cos\delta$ for case A$_2$ (normal spectrum).}
\end{center}
\end{figure}

\begin{figure}
\begin{center}
\includegraphics[scale=0.50,angle=-90]{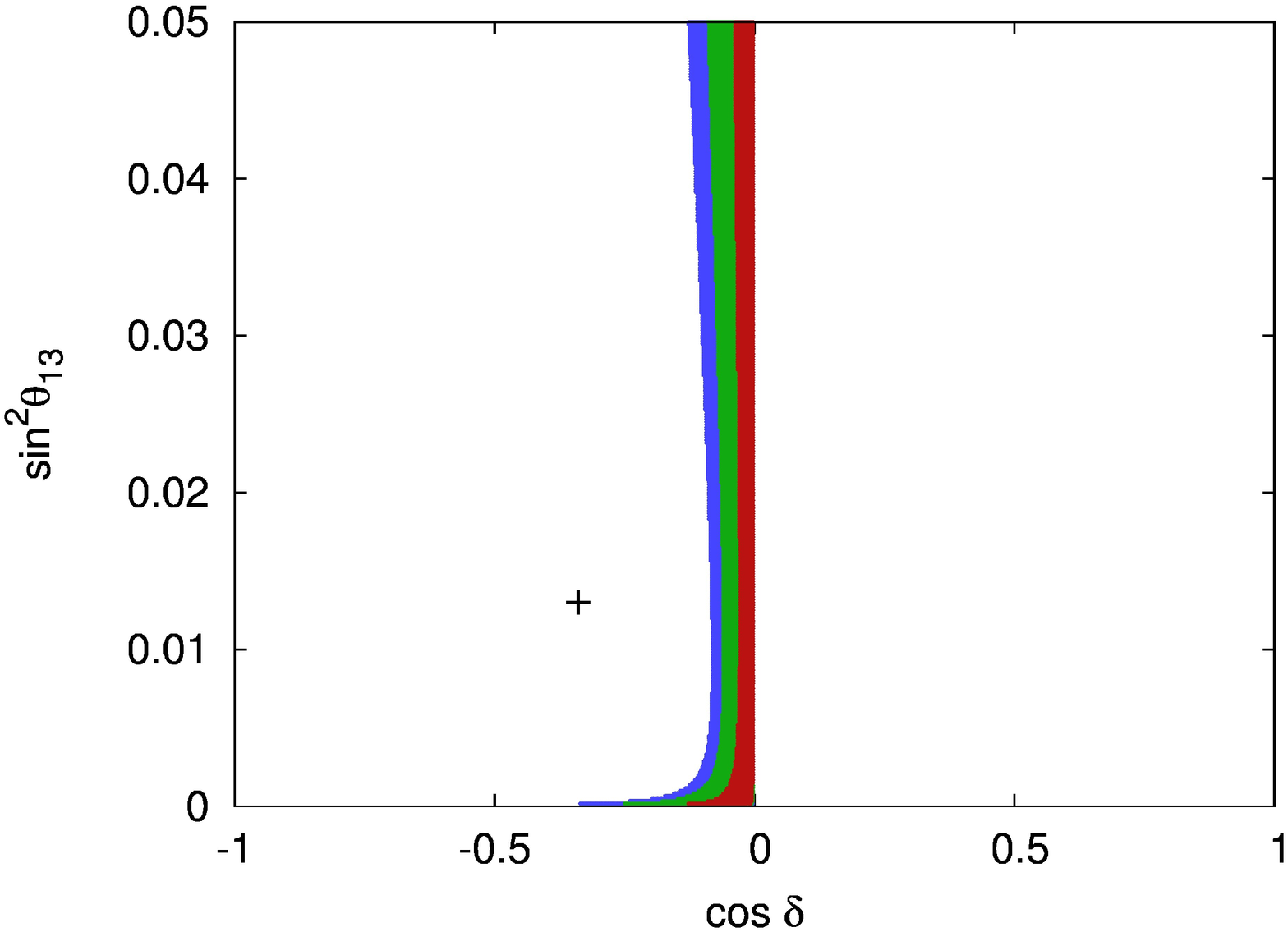}
\caption{The relation between $\sin^2\theta_{13}$ and $\cos\delta$ for case B$_1$ (normal spectrum).}
\end{center}
\end{figure}

\begin{figure}
\begin{center}
\includegraphics[scale=0.50,angle=-90]{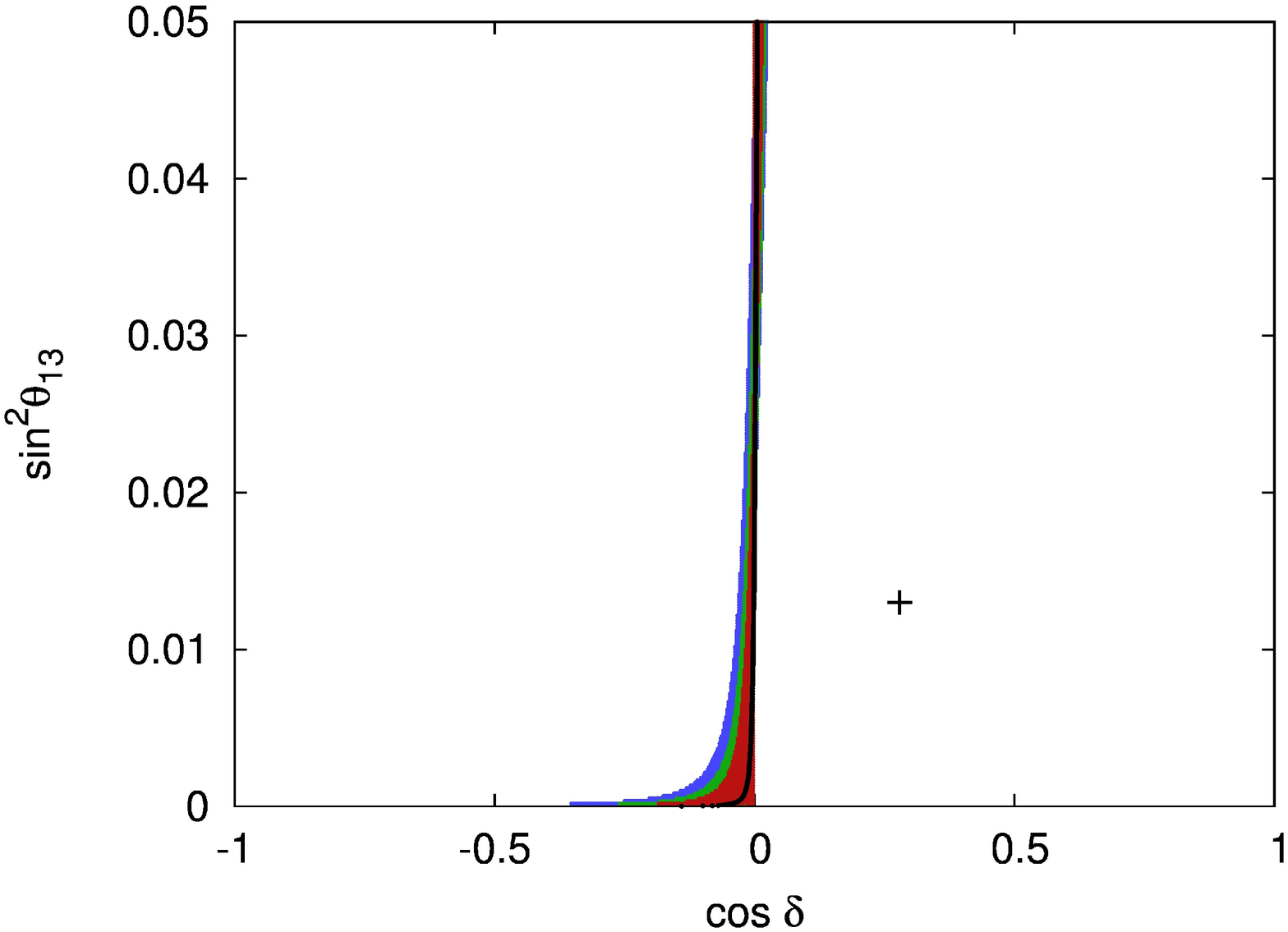}
\caption{The relation between $\sin^2\theta_{13}$ and $\cos\delta$ for case B$_1$ (inverted spectrum).}
\end{center}
\end{figure}

\begin{figure}
\begin{center}
\includegraphics[scale=0.50,angle=-90]{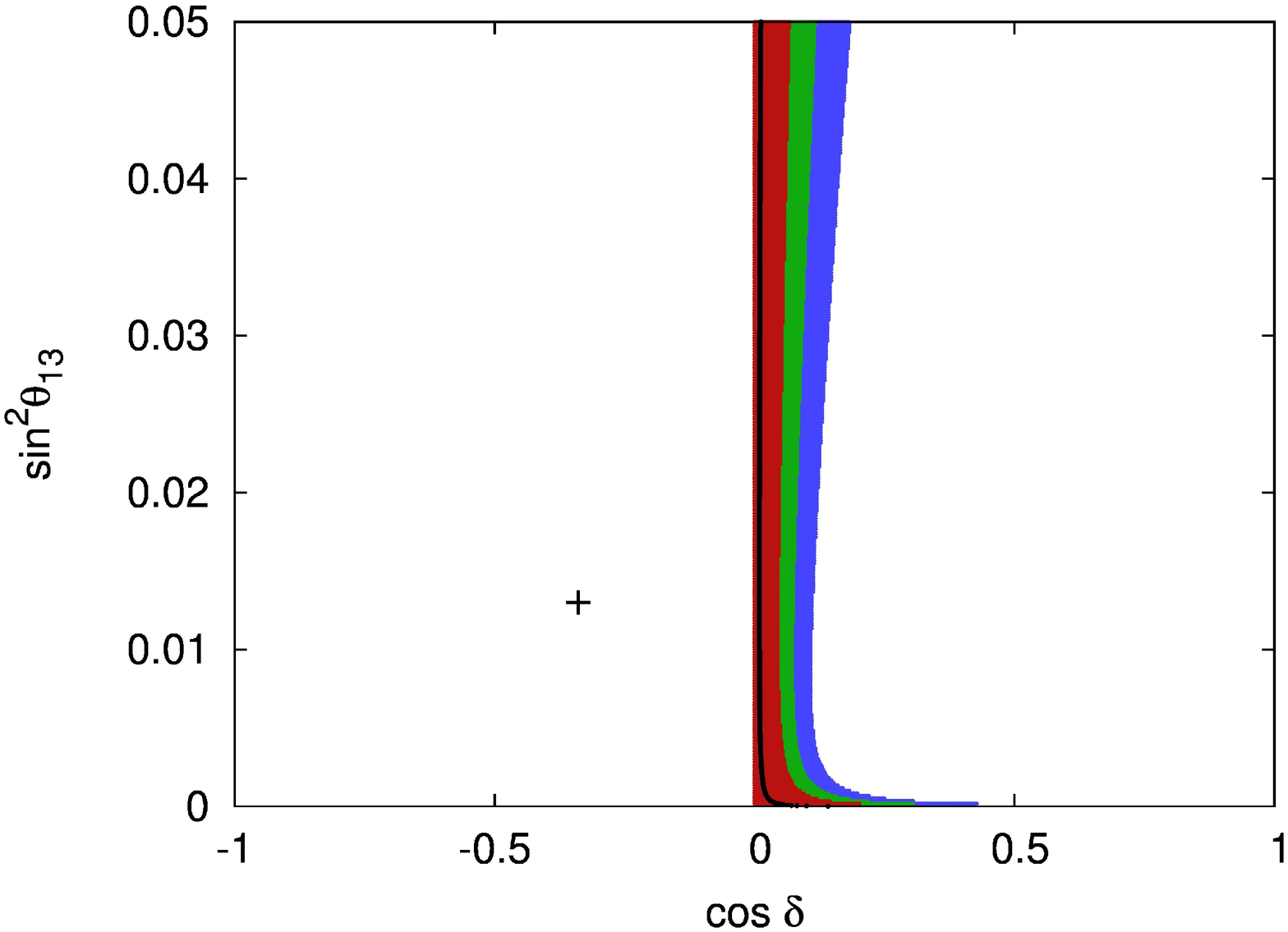}
\caption{The relation between $\sin^2\theta_{13}$ and $\cos\delta$ for case B$_2$ (normal spectrum).}
\end{center}
\end{figure}

\begin{figure}
\begin{center}
\includegraphics[scale=0.50,angle=-90]{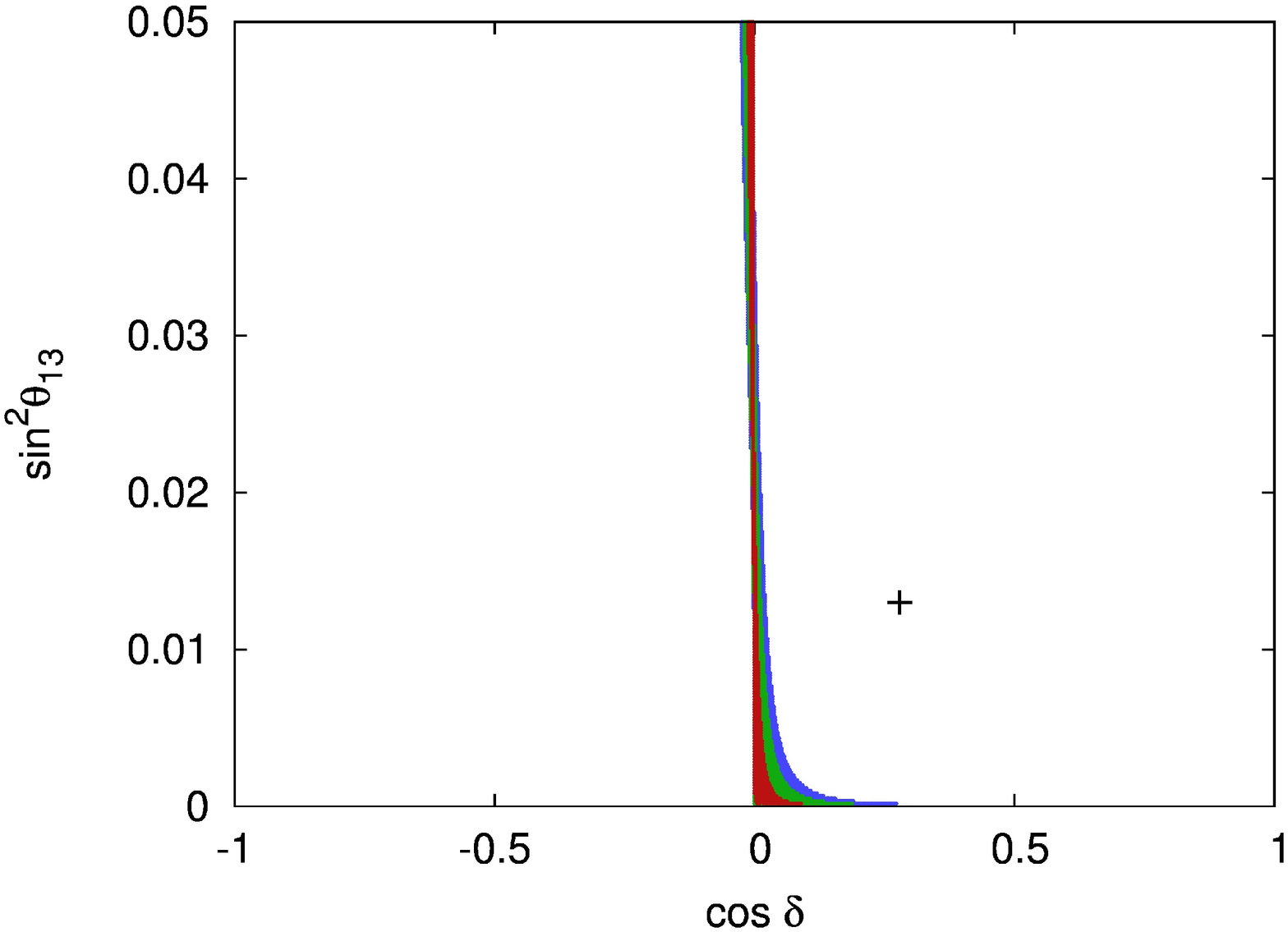}
\caption{The relation between $\sin^2\theta_{13}$ and $\cos\delta$ for case B$_2$ (inverted spectrum).}
\end{center}
\end{figure}

\begin{figure}
\begin{center}
\includegraphics[scale=0.50,angle=-90]{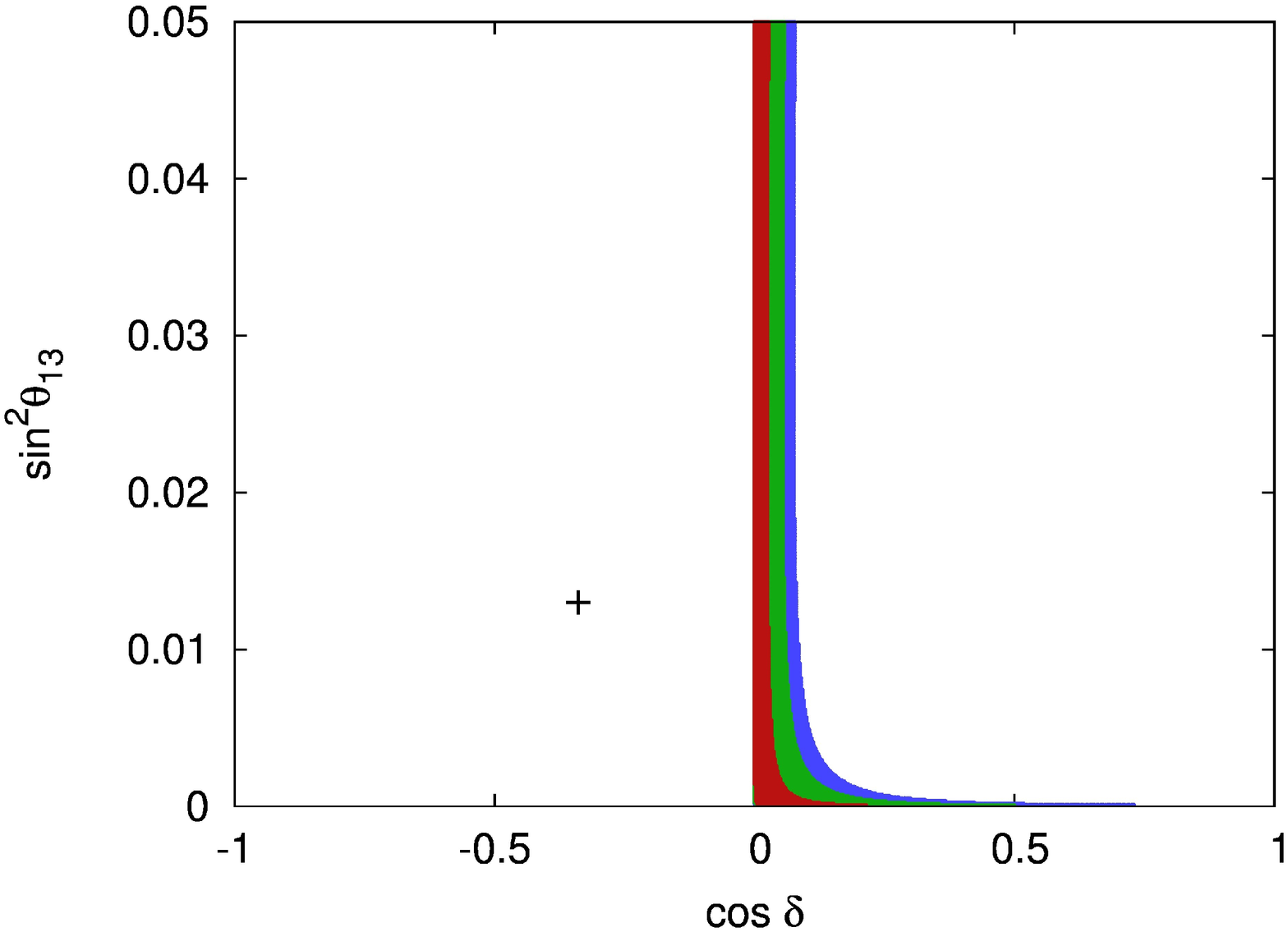}
\caption{The relation between $\sin^2\theta_{13}$ and $\cos\delta$ for case B$_3$ (normal spectrum).}
\end{center}
\end{figure}

\begin{figure}
\begin{center}
\includegraphics[scale=0.50,angle=-90]{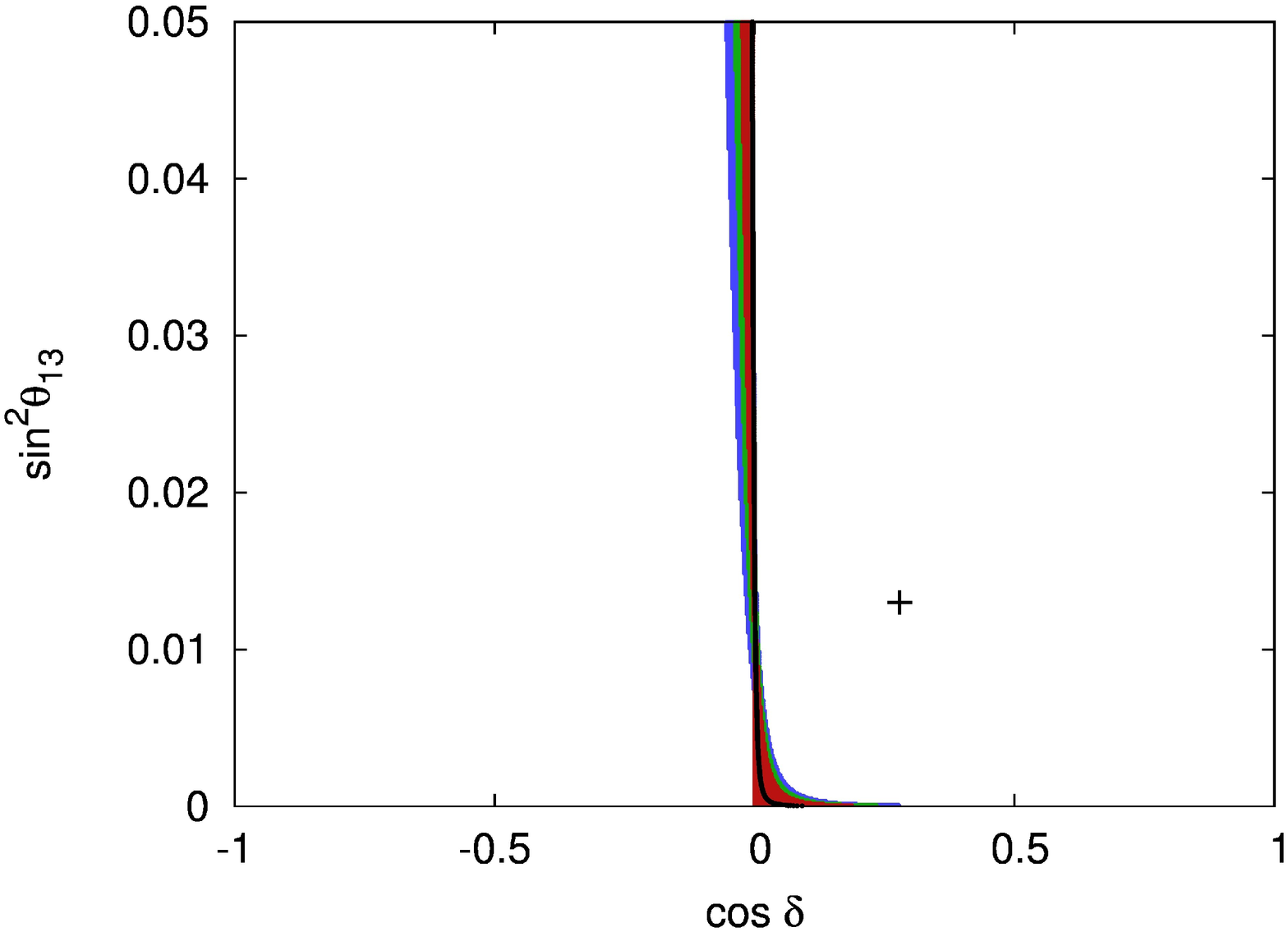}
\caption{The relation between $\sin^2\theta_{13}$ and $\cos\delta$ for case B$_3$ (inverted spectrum).}
\end{center}
\end{figure}

\begin{figure}
\begin{center}
\includegraphics[scale=0.50,angle=-90]{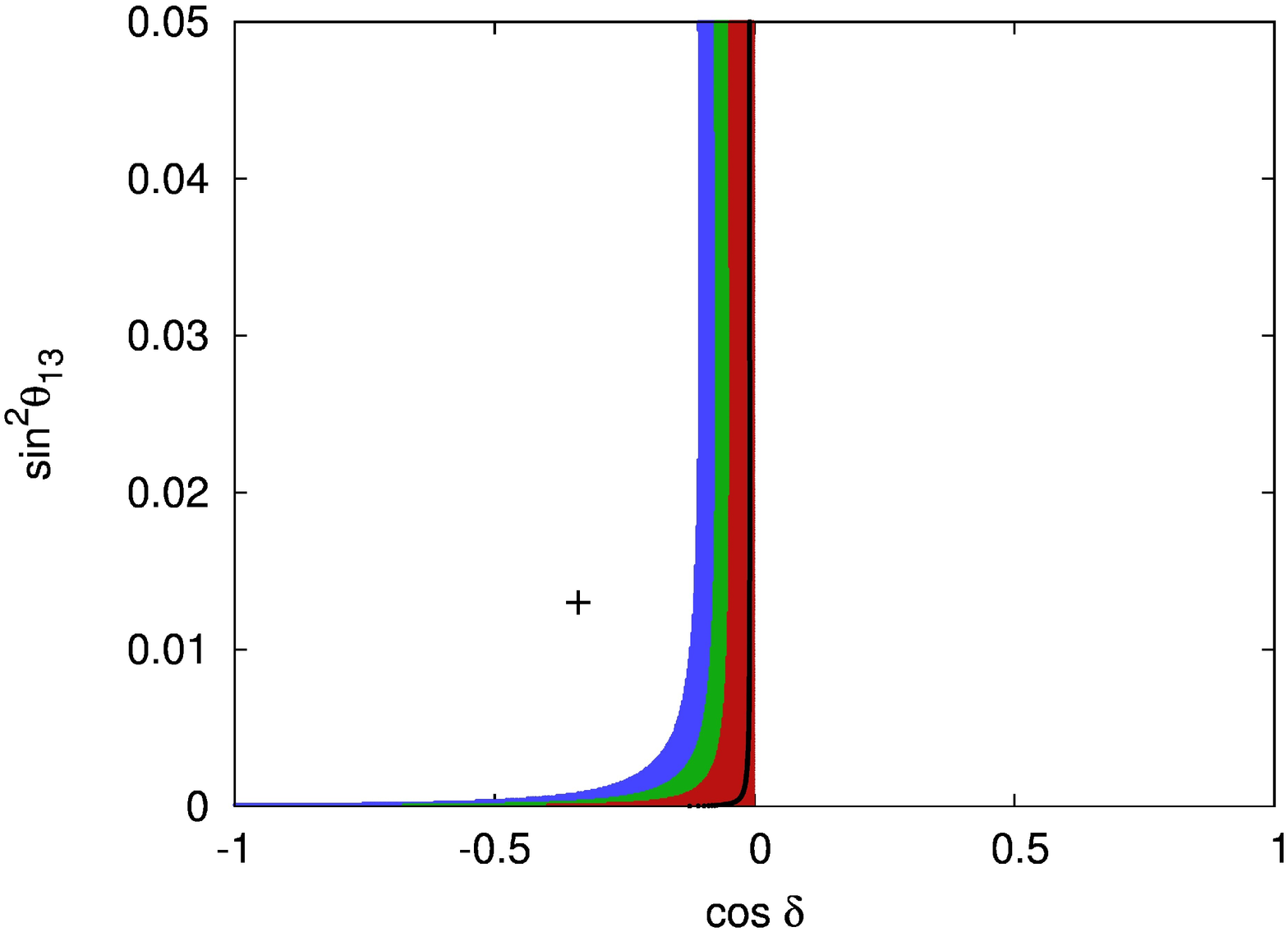}
\caption{The relation between $\sin^2\theta_{13}$ and $\cos\delta$ for case B$_4$ (normal spectrum).}
\end{center}
\end{figure}

\begin{figure}
\begin{center}
\includegraphics[scale=0.50,angle=-90]{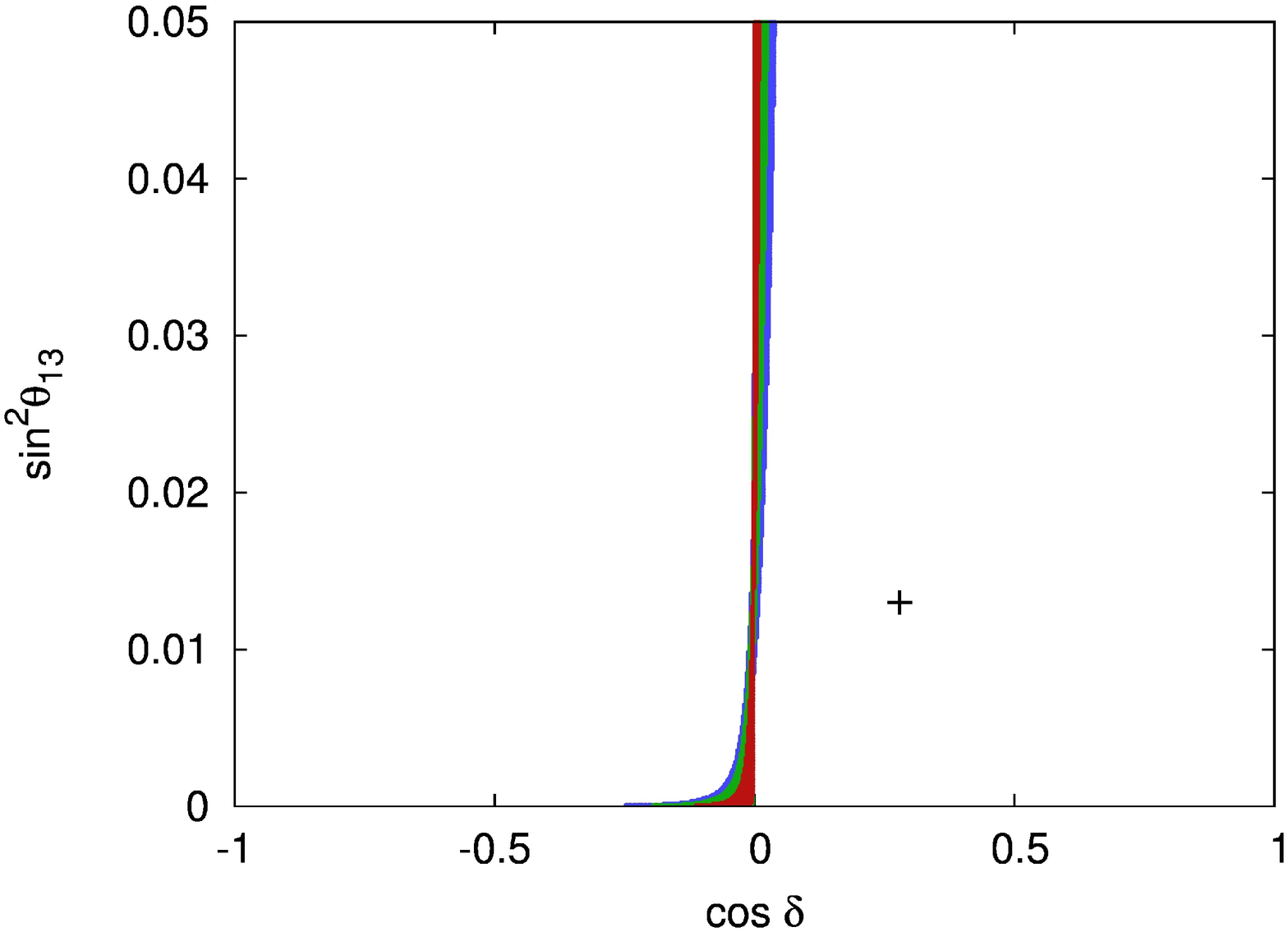}
\caption{The relation between $\sin^2\theta_{13}$ and $\cos\delta$ for case B$_4$ (inverted spectrum).}
\end{center}
\end{figure}

\section{Conclusions}

In the light of the recent T2K result which points towards a large reactor mixing angle $\theta_{13}$, we reconsidered the interesting case of two texture zeros in the neutrino mass matrix.
In particular we studied the correlation between the reactor mixing angle $\theta_{13}$ and the Dirac CP phase $\delta$
for the viable cases classified in~\cite{FGM} as A$_1$, A$_2$, B$_1$, B$_2$, B$_3$, B$_4$ and C.
All of these cases are still compatible with the global fit of the neutrino data at 3$\sigma$, but only the cases A$_1$ and A$_2$ predict the reactor angle to be different from zero at 3$\sigma$. In particular for the case A$_1$, asserting all the free parameters their best fit values, predicts $0.012\le\sin^2 \theta_{13}\le0.024$ while for the case A$_2$ assuming the best fit values predicts $0.014\le\sin^2 \theta_{13}\le0.032$.

\section*{Acknowledgments}

This work was supported by the Spanish MICINN under grants FPA2008-00319/FPA, FPA2011-22975 and MULTIDARK
CSD2009-00064 (Consolider-Ingenio 2010 Programme), by Prometeo/2009/091 (Generalitat Valenciana),
by the EU Network grant UNILHC PITN-GA-2009-237920. S. M. is supported by a Juan de la Cierva
contract. E. P. is supported by CONACyT (Mexico).

\end{document}